\documentclass[twocolumn,preprintnumbers,secnumarabic,amsmath,amssymb,superscriptaddress,nofootinbib,floatfix,prl]{revtex4}

\usepackage{graphicx}
\usepackage{dcolumn}
\usepackage{bm}

\def\beq{\begin{equation}}
\def\eeq{\end{equation}}
\def\be{\begin{equation}}
\def\ee{\end{equation}}
\def\bea{\begin{eqnarray}}
\def\eea{\end{eqnarray}}

\def\ab{{\bf a}}
\def\bb{{\bf b}}

\begin{document}

\preprint{IPPP/10/24, DCPT/10/48, Pi-cosmo-178}

\title{On detection of extra dimensions with gravitational 
waves from cosmic strings}

\medskip\
\author{Eimear O'Callaghan}%
\email[Email:]{e.e.o'callaghan@durham.ac.uk}
\affiliation{
Institute for Particle Physics Phenomenology, Department of Physics,
South Road, Durham, DH1 3LE, UK
}
\author{Sarah Chadburn}%
\email[Email:]{s.e.chadburn@durham.ac.uk}
\affiliation{
Centre for Particle Theory, Department of
Mathematical Sciences, 
South Road, Durham, DH1 3LE, UK
}
\author{Ghazal Geshnizjani}%
\email[Email:]{ggeshnizjani@perimeterinstitute.ca}
\affiliation{
Perimeter Institute for Theoretical Physics,
31 Caroline Street North, Waterloo ON, N2L 2Y, Canada
}
\author{Ruth Gregory}
\email[Email:]{r.a.w.gregory@durham.ac.uk}
\affiliation{
Institute for Particle Physics Phenomenology, Department of Physics,
South Road, Durham, DH1 3LE, UK
}
\affiliation{
Centre for Particle Theory, Department of
Mathematical Sciences, 
South Road, Durham, DH1 3LE, UK
}
\author{Ivonne Zavala}%
\email[Email:]{zavala@th.physik.uni-bonn.de}
\affiliation{
Bethe Center for Theoretical Physics and
Physikalisches Institut der Universit\"at Bonn,
Nu\ss allee 12, D-53115 Bonn, Germany
}

\date{\today}
\begin{abstract}
We show how the motion of cosmic superstrings in extra dimensions
can modify the gravitational wave signal from cusps.
Additional dimensions both round off cusps, as well
as reducing the probability of their formation, and thus give
a significant dimension dependent damping of the gravitational waves.
We look at the implication of this effect for LIGO and LISA,
as well as commenting on more general frequency bands.

\end{abstract}

\maketitle

The notion that nature might have extra dimensions has been with us for 
some time, but only recently have we revisited it with a view
to obtaining direct observational or experimental consequences.
The idea of Large Extra Dimensions (LED's), \cite{LED}, has given 
new possibilities both for compactifying nature's extra dimensions, 
as well as allowing a much richer gravitational phenomenology.
In particular, brane inflation \cite{braneinf} uses ideas from string 
theory, with inflation driven by the motion of a brane on some
stabilised internal manifold, \cite{KKLT}.
A key side effect of brane inflation is the formation of cosmic
strings, \cite{BCS,JST} as a by product of brane
annihilation (for reviews see \cite{CSSrev}),
and can have a wide range of physical parameters and properties.
The observation of such cosmic strings would therefore provide direct
evidence for string theory, as well as giving us valuable information
on inflation and the early universe. 

Cosmic strings, \cite{CS}, were originally popular as
an alternative to inflation, but were soon found to be inconsistent with
the emerging measurements of fluctuations
in the microwave background \cite{CMBrout}, although their
existence is not entirely ruled out \cite{CSfit}.
From the cosmological point of view, the internal structure of the cosmic 
string is irrelevant, and the string is taken to have zero width
with a Nambu action: $S = -\mu \int d^2 \sigma \sqrt{\gamma}$,
where $\mu$ is the mass per unit length of the string.
Together with rules for intercommutation \cite{IC}, or how 
crossing strings interact, this
gives the basic physics of how a network of cosmic strings will evolve.
Incorporating gravitational effects via a linearized approximation
indicates how fast energy is lost from the network, \cite{gwave}, and
putting all these pieces together gives the scaling
picture of the original cosmic string scenario \cite{NET}.

For cosmic superstrings the picture is similar, but
there are crucial differences. One is that the strings will
now not necessarily intercommute when they intersect \cite{ICSC}, a
simple way of understanding this is to imagine that the strings ``miss''
in the internal LED's. This clearly has a significant impact on one
of the drivers of network evolution, and leads to a denser network,
\cite{AS,JST}. 

Currently, gravitational wave experiments are most likely to
detect cosmic strings, with constraints on parameter space, \cite{obsv}, 
being derived using the Damour-Vilenkin (DV) results \cite{DV}. 
This calculation was performed in 4 spacetime dimensions,
however, while the reduced intercommutation probability
was taken into account, to our knowledge there has been
no systematic investigation of the impact of motion in the extra dimensions 
on the gravitational waves from cosmic strings.
In this letter we include these extra dynamical degrees
of freedom, and find a potentially significant moderation of 
the DV result, even when a phenomenologically
motivated cut-off is imposed. The basic physics behind the effect
is the extra degrees of freedom associated with the extra dimensions
which not only reduce the 
probability of cusp formation, but also round off the cusp producing
a narrowing of the gravity wave beam and hence a loss of power. 
The combination of these effects drops the gravitational wave event rate,
power, and hence detectable signal, thus altering current bounds 
\cite{obsv} from gravitational wave experiments.

To understand how this comes about, recall that a string obeying the
Nambu action sweeps out a worldsheet
in spacetime: $X^\mu=(t, \frac{1}{2}[\ab(\sigma_-)+\bb(\sigma_+)])$, 
where $\sigma_\pm = t \pm \sigma$ are the lightcone
variables on the worldsheet ($\sigma\in[0,L]$ for a closed
loop of length $L$). With the conventional gauge choice, 
$\ab'$ and $\bb'$ are constrained to lie on a unit
``Kibble-Turok'' sphere \cite{KT}.
At a cusp, $\ab' = \bb'$, and the left and right moving velocities
coincide, ${\dot X}^\mu_+ = {\dot X}^\mu_- = \ell^\mu = (1,{\bf n}')$;
the string instantaneously reaches the speed of light, and thus
there is a strong gravitational interaction. Cusps are therefore
transient but powerful events, and beam out a strong pulse of
gravitational radiation in a cone centered on the cusp.
As they are generic on string trajectories in 3 space dimensions 
(3d) this can lead to a potentially measurable gravitational signal.

In the seminal papers \cite{DV}, DV examined
gravitational radiation from cosmic strings, assessing for what
range of mass per unit length the string could potentially be visible
to the next generation of gravitational wave detectors. 
They first computed the amplitude of an individual cusp 
GWB as a function of the mass per unit length of the string,
obtaining the logarithmic cusp waveform:
\be
h^{\rm cusp}(f,\theta) \sim \frac{G\mu L^{2/3}}{r|f|^{1/3}} H[\theta_m-\theta]
\label{4Dwform}
\ee
where $f = \omega_m/2\pi = 2m/L$ is the frequency,
$H$ is the Heaviside step function, with $\theta$ the angle
between the wave vector ${\bf k}$ and the cusp vector ${\bf n}'
={\bf a}' = {\bf b}'$, and $\theta_m \simeq (2/Lf)^{1/3}$
a cut-off giving the opening angle of the cone in which the
GWB beams out from the cusp.

\begin{figure}[htbp!]
\centering
\includegraphics[width=0.45\textwidth]{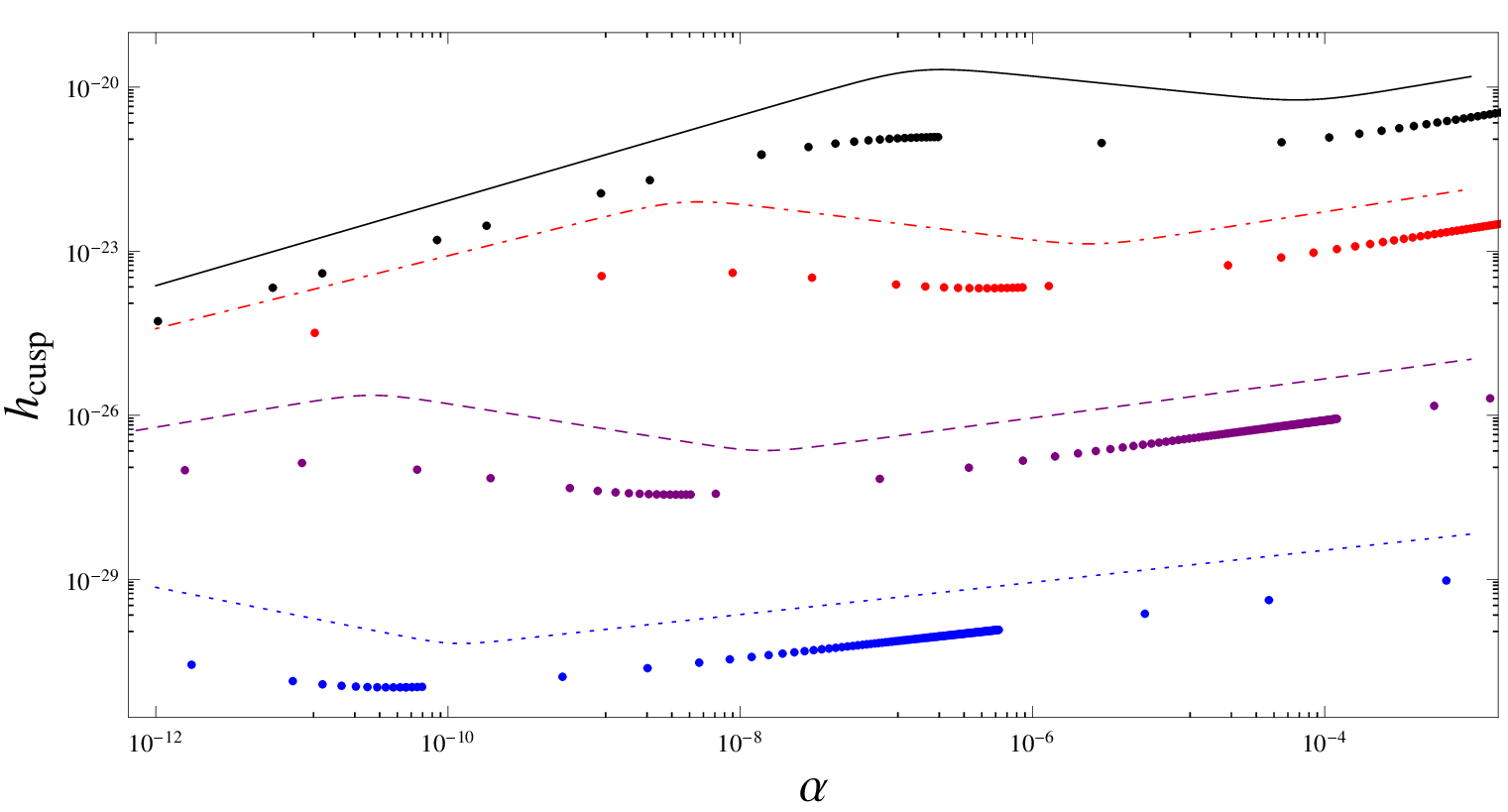}
\caption{
A direct comparison with the DV plot \cite{DV}, showing the
GWB amplitude at $f = 150$ Hz as a function of $\alpha$. 
Solid lines show the interpolating function result, the dots 
correspond to exact numerical results. 
From top to bottom the plots are: 3d DV in black,
in red (dot-dash) $n=1$, purple (dashed) $n=3$, and blue (dotted) $n=6$.
}
\label{fig:LIGO}
\end{figure}

\begin{figure}[htbp!]
\centering
\includegraphics[width=0.45\textwidth]{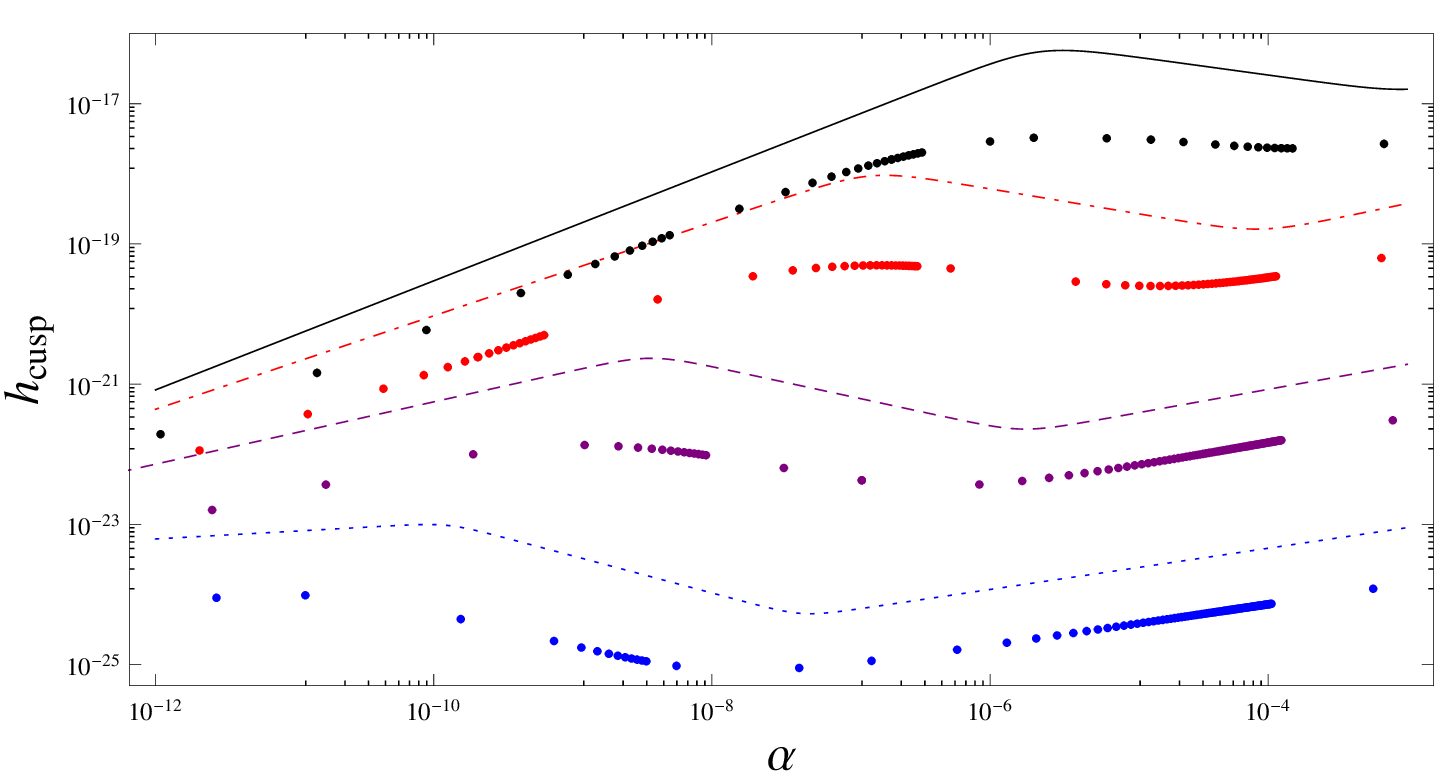}
\caption{
As figure \ref{fig:LIGO}, but with $f = 3.9$ mHz appropriate to
the LISA detector.
}
\label{fig:LISA}
\end{figure}

In an expanding universe, the waveform frequency is redshifted in 
the obvious way, $f\to(1+z)f$, and $r$ in the asymptotic waveform 
must be replaced by the physical distance, $a_0 r = (1+z) D_A(z)$,
where $D_A(z)$ is the angular diameter distance at redshift $z$.
To find the background for a cosmological network of strings,
DV used the one scale model, $L \sim \alpha t$,
$ n_L(t) \sim 1/ (\alpha t^3) $, to write the loop length and network
density in terms of cosmological time. ($\alpha \sim 50 G\mu$ 
is a constant representing the rate of energy loss from string loops
\cite{gwave}).
The expected number of cusp events per unit spacetime volume is
then given by $\nu(z) \sim \,{\cal C}n_L/PT_L \sim 2{\cal C}/P\alpha^2 t^4$,
where ${\cal C}$ is the average number of cusps per loop period $T_L 
= L/2 \sim \alpha t/2$, and $P$ is the intercommutation probability, 
\cite{ICSC}, which DV take in the range $10^{-1}-10^{-3}$.
From this they obtain an estimate of the rate of GWB's per unit 
spacetime volume at redshift $z$ as
\be
d{\dot N} \sim \frac{\nu(z)}{(1+z)} \frac{\pi \theta_m^2 (z)
D_A(z)^2}{(1+z)H(z)} dz \; .
\label{dndot}
\ee
The final step of the DV argument is to integrate out until
a desired event rate at an experimentally motivated
fiducial frequency is obtained, then invert to find the redshift 
which dominates the signal. Evaluating the gravitational 
wave at this redshift and frequency then gives the amplitude. 
In practise, DV use interpolating functions for the angular diameter 
and cosmological time, which allows them to approximate these expressions 
analytically, and obtain a direct form of the amplitude (the black lines
in figures \ref{fig:LIGO}, \ref{fig:LISA}).

With extra dimensions, 
the motion of the string in the internal dimensions causes it to
appear to slow down in our noncompact space dimensions, which
allows the left and right moving modes to misalign in
momentum space, thus avoiding an exact cusp, which becomes
a highly special feature in higher space dimensions.
We need to generalize the notion of a ``cusp'', and
estimate its probability. A near cusp event (NCE) is 
a local minimum of $|{\bf a}' - {\bf b}'| = 2\Delta \ll1$, and is
parametrized by $\Delta$, which measures how close 
to an exact cusp (EC) this event is.  Assuming a uniform distribution
of solutions in parameter space, and modelling simple higher
dimensional loop solutions (see \cite{compan} for full details and toy
loop solutions), we find that the number of NCE's with 
$|\ab' - \bb'|_{\rm min} \leq 2\Delta$ in a generic loop is ${\cal N}
(\Delta) \simeq  \Delta^n$ (since all loops have $|\ab' - \bb'| 
\leq 2$ at all points on their trajectory).

We now compute the waveform for a NCE. Since these strings are
formed in brane inflation scenarios, the flux
stabilization procedure that prevents dangerous cosmological
moduli evolution, \cite{KKLT}, should also prevent the excitation of
internal KK degrees of freedom. Thus, we can use the standard Einstein 
propagator in calculating the gravitational radiation from a cusp.
The main difference between the EC and the NCE is that the
4-velocity ${\dot X}^\mu = (1, (\ab'+\bb')/2)$ need not be
null, and that the individual left and right moving
velocities need not be aligned. The effect of this misalignment is
similar to the misalignment between the cusp direction vector
and the gravitational wave vector, and performing the computation in detail,
\cite{compan}, shows that the the waveform
of the NCE is the same as (\ref{4Dwform}), with the proviso that the
cone opening angle in is decreased to
$\theta_\Delta = \theta_m - \Delta$.

Cosmologically, a general network will have a
range of NCE's with different $\Delta$ values, up to and including
the cutoff value when the GWB beaming cone closes off.
We must therefore calculate the GWB event rate, ${\dot N}$, as
a function of $\Delta$, replacing the solid angle $\theta_m^2(z)$ 
by $(\theta_m(z)-\Delta)^2$, and $\nu(z) \to \nu(z,\Delta)
= {\cal C}(\Delta) n_L/PT_L$, 
where ${\cal C}(\Delta)$ is the local probability density of NCE's
for the network. Assuming that the loops are spread evenly 
in the parameter space of solutions, 
${\cal C}(\Delta) = {\cal N}'(\Delta) = n\Delta^{n-1}$,
and we integrate over $\Delta$ to obtain the nett
effect of {\it all} possible NCE's:
\be
\frac{d{\dot N}_{\rm NCE}}{dz}
= \frac{2 \theta_m(z)^{n+2}}{(n+1)(n+2)}
\frac{n_L(z)}{PT_L(z)} \frac{\pi D_A(z)^2}{(1+z)^2 H(z)}
\label{newrate}
\ee

Figures \ref{fig:LIGO} and \ref{fig:LISA} show the gravitational wave 
amplitude for the LIGO and LISA detectors respectively. In each case, we
used the same fiducial frequencies as DV: the waveform has the same
profile hence the signal to noise analysis remains the same.
For direct comparison we used the interpolating function method,
but also include an exact numerical computation for the
concordance cosmology ($\Omega_r  = 4.6 \times 10^{-5}, \Omega_m 
= 0.28, \Omega_\Lambda =1-\Omega_m-\Omega_r$).

Clearly, the motion in the extra dimensions 
has a significant effect on the GWB amplitude, however, to what extent
is this result a feature of our assumptions? The basic reason
for the suppression of the signal is the distribution
over the near cusp parameter $\Delta$. This was derived assuming
a uniform distribution in solution space, and a zero width string.
Let us deal with each in turn.

One objection to having a uniform distribution in solution space is the 
notion that compact extra dimensions must somehow constrain the allowed
parameter space of the string. Since cosmic strings form from 
the collision of a brane and anti-brane, it seems likely that 
they have significant initial momentum in the extra dimensions, 
thus it seems reasonable not to curtail solution space in this way.
However, one might worry that if the loop
wraps back and forth across the extra dimension(s) the string
has more opportunity to self intersect, and that this will
result in a restriction on parameter space.
We modelled this, \cite{compan}, by exploring the self intersection of
a 4d family of loops with a 3d limit. In 3d, about 30\% of the parameter
space had self intersections, but in 4d, once again the measure of
solution space with self intersections became zero by a similar
parametric argument as for the cusp. 

\begin{figure}[htbp!]
\centering
\includegraphics[width=0.45\textwidth]{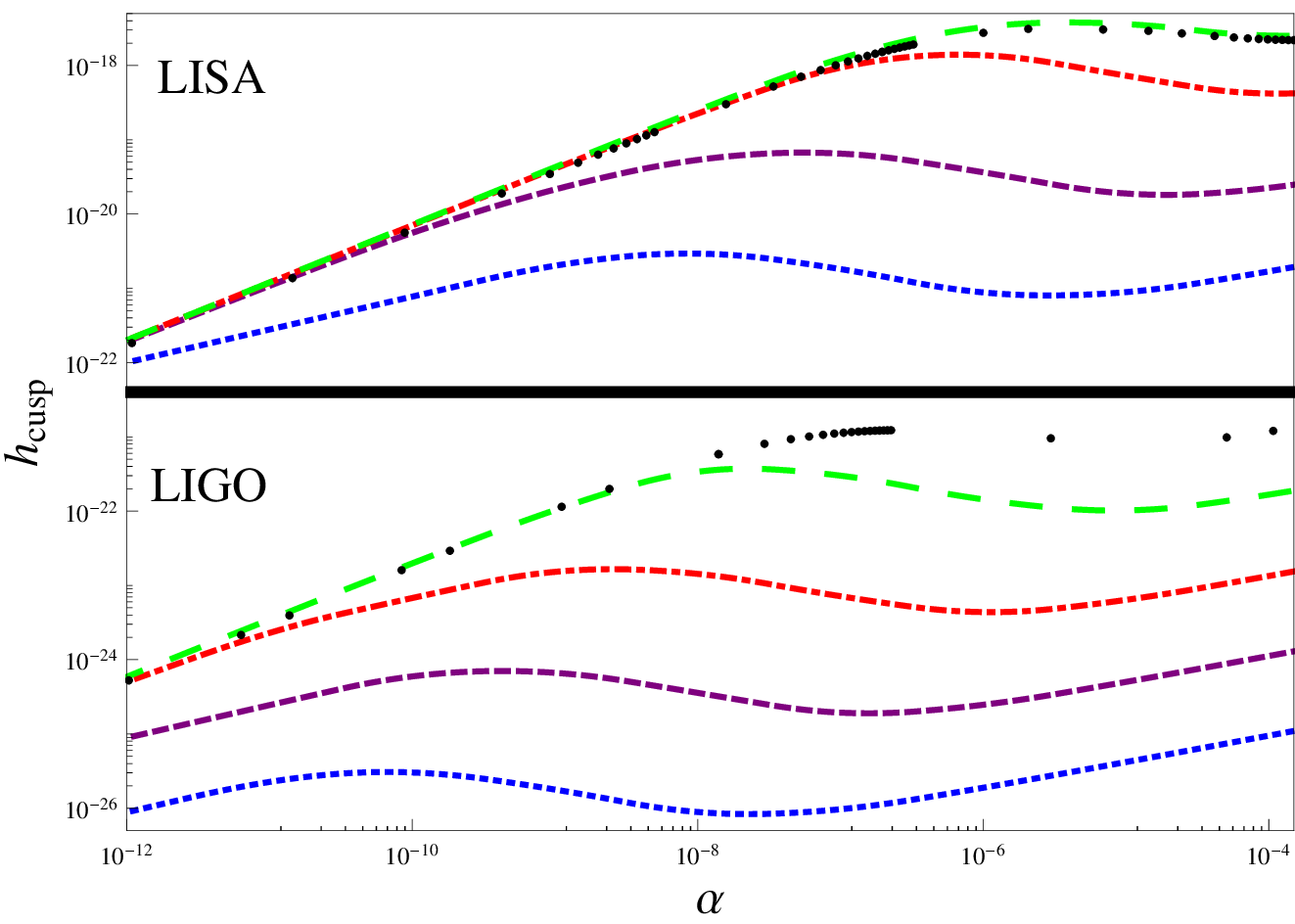}
\caption{
A plot of the amplitude for LISA and LIGO with the modified measure 
(\ref{deltaint0}) and $P=10^{-3}$, varying $\Delta_0$ with fixed $n=3$.
From top to bottom: the 3d DV result, then $\Delta_0 = 10^{-4},
10^{-3}, 10^{-2}$, and $0.1$ respectively.
}
\label{fig:delta0}
\end{figure}

The clear outcome of testing exact loop trajectories is that for a zero
width string, there is no restriction on parameter space from compact
extra dimensions. However, cosmic strings have finite width, $w$,
and while this is smaller than the internal LED size, $R$, we would expect 
the ratio $w/R$ to enter into the parametric computation. We model this 
by restricting $\Delta\in[0,\Delta_0]$ with $\Delta_0$ related to $w/R$, 
and normalize ${\cal C}$ so that $ {\cal N}(\Delta_0) = 1$, 
i.e.\ ${\cal C}(\Delta) = n\Delta^{n-1}/\Delta_0^n$. This modifies
the dependence of $d{\dot N}_{\rm NCE}/dz$ on $\theta_m$ to
\bea
\int\limits_0^{{\rm min}\{\Delta_0,\theta_m\}} \hskip -12mm
&&{\cal C}(\Delta) \left ( \theta_m(z) - \Delta \right )^2
= \frac{2 \theta_m(z)^{n+2} H[\Delta_0 - \theta_m] }
{\Delta_0^n(n+1)(n+2)} \nonumber \\
&+& \left (\theta_m^2 (z) - \frac{2n\Delta_0\theta_m (z)}{n+1} 
+ \frac{n\Delta_0^2}{n+2} \right ) H[\theta_m-\Delta_0] \;. \hskip 9mm
\label{deltaint0}
\eea
To test this alternate expression, we took values of $\Delta_0 
= 0.1 - 10^{-4}$ (see figure \ref{fig:delta0}).
From (\ref{deltaint0}), we see that the
effect of $\Delta_0$ is to shift the behaviour from (\ref{newrate}) for 
$\theta_m < \Delta_0$, towards a $\theta_m(z)^2$ form as $\theta_m$
grows. For $\theta_m > 50 \Delta_0$, the 3d result is recovered.
Since $\theta_m(z) \propto (G\mu)^{-1/3}$, the results converge to 
the 3d value at larger $\Delta_0$ for smaller $G\mu$.
\begin{figure}[htbp!]
\centering
\includegraphics[width=0.45\textwidth]{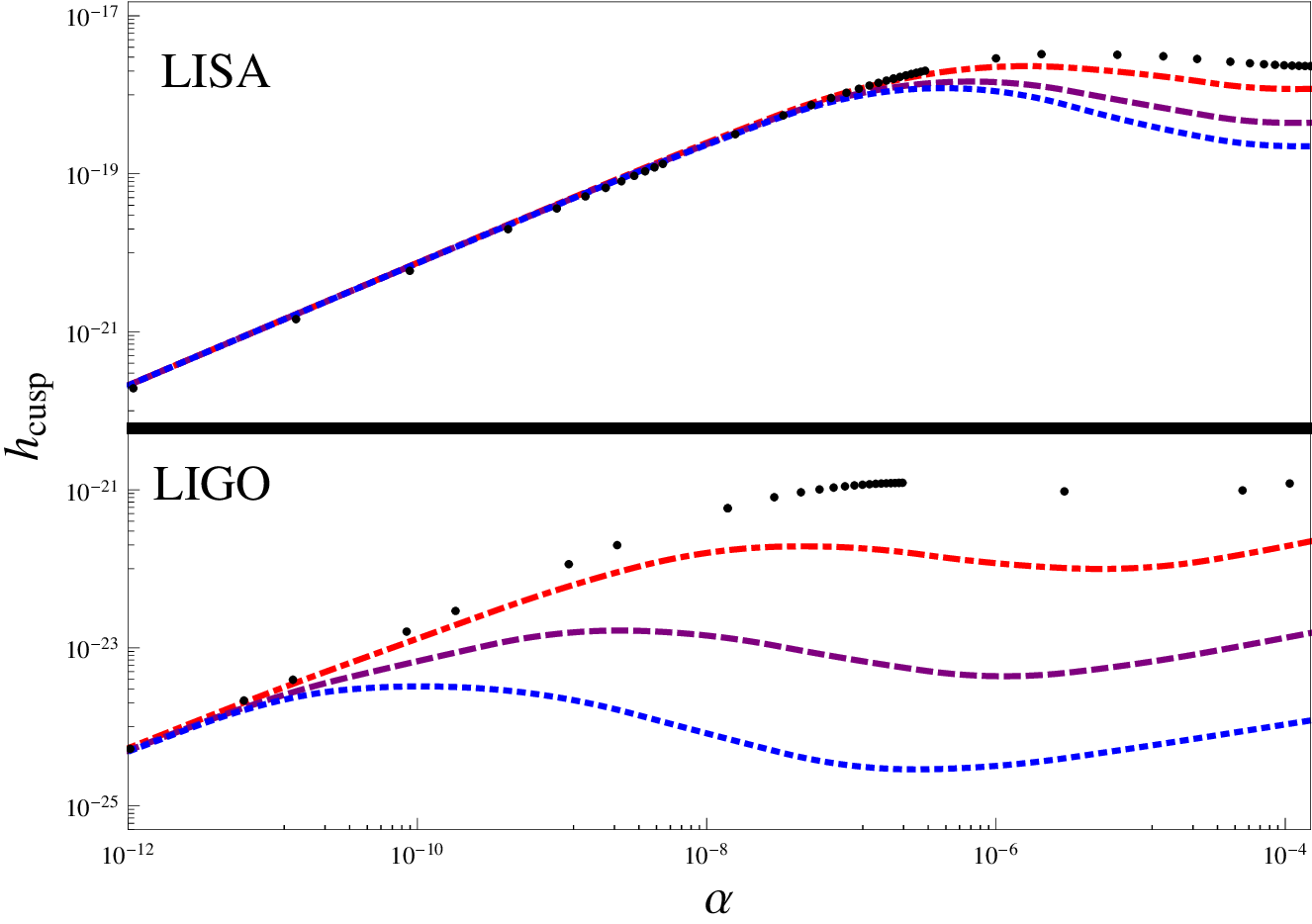}
\caption{
A plot of the amplitude for LISA and LIGO with the modified measure 
(\ref{deltaint0}) and $P=10^{-3}$, fixing $\Delta_0 = 10^{-3}$, 
and varying $n$.  From top to bottom: the 3d result,
and the extra dimension plots with $n=1, 3$, and $6$ respectively.
}
\label{fig:varyn}
\end{figure}

To sum up: We have studied the impact of motion in extra dimensions
on the GWB signal from cusp events on cosmic string loops. 
We find a potentially significant moderation of the signal, even after
taking into account finite width effects and the size of the extra
dimension. Clearly further work is required to get better control
of the approximations being used, in particular to take into account
more complex compactification geometries, however it does seem that
motion in internal dimensions is important. 
Although we have focussed on LIGO and LISA,
we should also comment on alternative GW detectors.
It is not difficult to see from the interpolating function approximation
that the higher the frequency, the greater the damping on the signal
due to the extra dimensions. Coupled with the inherent damping of
the signal at higher dimensions, we conclude that high frequency 
GW detectors, such as the electromagnetic detector of Cruise et
al.\ \cite{CRUISE}, have little hope of seeing this signal.
On the other hand, at low frequencies, such as those probed
by the pulsar timing array \cite{PULSAR}, the damping due to
extra dimensions is diminished. For the pulsar limits, the confusion noise,
as defined by DV, will be similar for both cosmic superstrings
and strings, since at such low frequencies we are at the limit of
the approximation used in the calculation of the waveform, and
the GWB cone has completely opened out.
This would suggest that we can no longer trust the waveform
(\ref{4Dwform}) or its extra dimensional analog that we have used throughout.
Finally, from the dependence of the signal on $n$,
the possibility arises that a positive detection of gravitational
radiation would not only confirm the general brane inflation scenario,
but could provide a means of determining the number of (effective)
extra dimensions.

\begin{acknowledgments}
We would like to thank A.Avgoustidis, J.J.Blanco-Pillado, D.Chung, L.Leblond,
S.Rajamanoharan, P.Shellard, G.Shiu, and M.Wyman for helpful discussions.
RG and IZ would like to thank the Perimeter Institute for hospitality
while this work was being undertaken, and RG would like to thank the
Cambridge CTC for hospitality while this work was being completed.
SC is supported by EPSRC, EOC is supported by EC FP6
through the Marie Curie EST project MEST-CT-2005-021074,
GG is suppported by the Government of Canada through Industry Canada and
by the Province of Ontario through the Ministry of Research, RG is
partially supported by STFC, and IZ is partially supported by the EC FP6
program MRTN-CT-2006-035863 and by the DFG SFB-Transregio 33.

\end{acknowledgments}


\end{document}